\documentclass[12pt,preprint,a4paper]{aastex}

\usepackage{graphics,epsf}
\usepackage{amsmath}                
\usepackage{amsfonts}               
\usepackage{amssymb}                
\usepackage{epsfig}                 
\usepackage{lscape}

\def \s{~\rm{s}}
\def \km{~\rm{km}}

\def \AU{~\rm{AU}}

\def \yr{~\rm{yr}}

\def \days{~\rm{day}}

\begin{document}

\title{FIRST VERSUS SECOND GENERATION PLANET FORMATION IN POST COMMON ENVELOPE BINARY (PCEB) PLANETARY SYSTEMS}

\author{Ealeal Bear\altaffilmark{1} and Noam Soker\altaffilmark{1}}

\altaffiltext{1}{Department of Physics, Technion - Israel Institute of
Technology, Haifa 32000 Israel; ealealbh@gmail.com; soker@physics.technion.ac.il.}

\begin{abstract}
We examine planets orbiting post-common envelope binaries (PCEBs) from the perspective of angular momentum evolution, and conclude that
the planets are more likely to be first generation (FG) planets than second generation (SG) planets.
FG planets were born together with the parent stars, while SG planets form later from a SG proto-planetary disk formed by mass-loss from the evolved
primary star during its red giant branch (RGB) phase or asymptotic giant branch (AGB) phase.
We find that in some systems the SG scenario requires that more than twenty percent of the SG proto-planetary disk mass ends in planets.
Although we cannot rule out SG planet formation in these systems, this fraction of mass that ends in planets is much higher than the value
commonly used in planet formation theories.
On the other hand, we find that for each of the systems we can build a progenitor system composed of a main-sequence binary system orbited by the appropriate planets.
This can be done if the secondary star was in a resonance with the inner planet.
To account for the progenitor properties we suggest that in cases where the secondary star has a mass of $\sim 0.1-0.2 M_\sun$,
it was formed in the same way planets are formed, i.e., from a disk.
\end{abstract}

\section{INTRODUCTION}
\label{sec:intro}

Planets have been inferred indirectly by eclipse timing variations around post common envelope binaries (PCEBs) where the secondary is a main sequence (MS) star orbiting a compact object on a close orbit.
The compact object is the core remnant of either an asymptotic giant branch (AGB) star, where it is a CO white dwarf (WD) or a NeMgO WD,  or the remnant
of a red giant branch (RGB) star.
In the later case the compact object might be a He WD, or a horizontal branch (HB) star, i.e., an sdB or sdO star, or a CO WD remnant of a HB star.
We refer to all the objects that are RGB-descendant stars as RGBD objects.
The evolution of PCEB hosting planets is still an open area, mainly due to open questions in the common envelope (CE) evolution itself (e.g., \citealt{Webbink1984}, \citealt{Nelemans2005}, \citealt{Ivanovaetal2013}, \citealt{Soker2014}).

Planets orbiting PCEBs can form together with the binary, i.e. first generation (FG) planets, or can form after the CE phase, i.e. second generation (SG) planets
\citep{Parsons2010b,Perets2010, Perets2011, Tutukov2012, PortegiesZwart2013, Schleicher2014, Volschow2014}.
Exoplanet systems around MS binary systems with a low-mass secondary star suggest that the FG scenario is plausible.
Examples include the planet Kepler-16b that orbits with a period of $228\days$ a stellar binary system of masses $M_1=0.69M_\odot$ and $ M_2=0.2M_\odot$ and
a period of $41.1 \days$ \citep{Doyle2011}, and Kepler-38b with a planet period of $105.6\days$ around a stellar binary systems having masses of
$M_1=0.95M_\odot$ and $ M_2=0.25M_\odot$ and a period of $18.8 \days$ (\citealt{Doyle2011,Orosz2012, Meschiari2013, Pelupessy2013}).
These systems are more compact than the progenitor systems we will study here, but show that planetary systems around stellar binary systems exit.

The idea of a SG planet formation goes back to the first exoplanet detected around a pulsar PSRB1257+12 \citep{WolszczanFrail1992}.
In this scenario the planet was formed from the material that was ejected in the supernova and formed a disk around the neutron star remnant \citep{Tavani1992}.
Formation of planets in a circum-PCEB disk is currently not understood well.
The existence of circumbinary disks (e.g., \citealt{Perets2010} and references therein), although around much wider binaries than
PCEB, hints that such disk can be formed and might lead to planet formation.
Another factor that can help SG planet formation is relevant in the case of AGB post-CE planetary systems, as the ejected material has high metallically,
favoring planet formation under the core accretion scenario \citep{Zorotovic2013} .

\cite{Schleicher2014} studied 12 PCEBs, elaborating on NN~Ser. They find indications for the presence of two different planet populations, which may reflect different formation mechanisms, including SG planets.
They give a detailed review on the SG planet formation mechanism (which we will not repeat here;
see also \citealt{Perets2010}), which they found to fit better to the properties of the 12 PCEB systems they have studied.
\cite{Schleicher2014} refer to the work of \cite{KashiSoker2011} on disk formation in CE evolution. We note however that the post-CE disk mentioned by \cite{KashiSoker2011}
is much smaller than the proto-planetary disk required in the SG scenario for these systems.
\cite{Mustill2013} and \cite{Horner2012a} examined the dynamical stability of NN~Ser and argue that the FG scenario is unlikely, as it is very hard to
find a progenitor planetary system for NN~Ser that was stable along the evolution. They prefer the SG scenario for NN~Ser.
We do note thought that for a very small fraction of initial conditions they do get long-live systems, longer than the main sequence life
time of the primary star (e.g., the two white squares in the left part of figure 3 of \citealt{Mustill2013}).
We will attribute these rare cases to systems where the binary system is in a resonance with the inner planet.

Our interpretation of the literature on the SG formation scenario is that the short timescale available for planet formation supports the disk instability planet formation which is not free of problems either (e.g., \citealt{Matsuo2007,Janson2012}).
\cite{Zorotovic2013} pointed out that planets will be more likely to form around PCEBs with high mass CO WDs than around low-mass He WDs,
as the former are descendants of AGB stars where the envelope had suffered metal enrichment. Many of the PCEB planets however, are descendants of
RGB stars.
\cite{Zorotovic2013} noted that the cooling age of the WD in NN~Ser is on the order of a million years \citep{Parsons2010a}, which
strongly constrains the SG process if planets form from planetsimals. Planet formation by gravitational instabilities also constrains the SG,
as the initial disk mass is not much larger than the final mass in planets.

Motivated by the recent works of \cite{Zorotovic2013} and \cite{Schleicher2014} we examine the same twelve systems
that \cite{Schleicher2014} have studied,
but we concentrate on angular momentum considerations.
In section \ref{sec:SG} we review the SG scenario in light of angular momentum evolution,  and in section \ref{sec:FG}
we review some constrains on the FG scenario.
{{We point out that the criteria used to examine the SG and FG scenario are not identical.
We confront each scenario with basic physical processes that challenge the scenario.
For the SG scenario the main one is the source of the angular momentum of the planetary system, while for the FG scenario the challenge is to
construct a circumbinary planetary system that can be the progenitor of the observed system.}}
Our summary and conclusions are in section \ref{sec:Summary}.

\section{EXAMINING THE SECOND GENERATION SCENARIO}
\label{sec:SG}
{{Readers interested in our main conclusion can skip section \ref{sec:SGdata} were we describe the data summarized in the Table 1,
and section \ref{sec:SGparameters} were we define the efficiency parameters, and go directly to section \ref{sec:SGconclusion} were we conclude our constrains on the SG scenario. }
\subsection{The PCEB and planetary systems}
\label{sec:SGdata}

The post-common envelope binary (PCEB) and the planetary systems studied are taken from \cite{Zorotovic2013} and \cite{Schleicher2014},
and are summarized in Table 1. The first nine columns after the name of each system list observational quantities:
(1) primary stellar mass; (2) secondary stellar mass; (3) binary orbital separation;
(4, 6, 8) mass, orbital semi-major axis, and eccentricity, of the first planet (not necessarily the inner one); (5 ,7, 9) the same for the second planet if exists.
The last 5 columns are quantities defined and calculated here: (10) the minimum angular momentum of the planetary system; (11) the initial progenitor mass estimated from an initial-final mass relation (for details see Table 1 remark $*_3$).
(12) the initial angular momentum efficiency factor; (13) the specific angular momentum efficiency factor; (14) our assessment on the likelihood of the SG scenario.
\renewcommand{\tabcolsep}{1pt}
\begin{table}
Table 1: Systems - SG Scenario
\newline
\footnotesize{
\begin{tabular}{|l|c|c|c|c|c|c|c|c|c||c|c|c|c|c|c|c|}
\hline
\small Column & 1 & 2& 3 & 4 & 5 &6& 7& 8& 9&10& 11 &12&13 & 14 \\
\hline
\small Name & $M_1$ & $M_2$  & $a_2$ & $m_{1p}^{*_1}$ & $m_{2p}^{*_1}$ &$a_{1p}$& $a_{2p}$& $e_{1p}$& $e_{2p}$&$J_{pm}$& $M_{1,0}^{*_3}$ &$\eta_{J2}^{*_4}$ & $\eta_{L2}^{*_4}$&SG \\
\hline
Units & $M_\odot$ & $M_\odot$ & $R_\odot$ & $M_{\rm J}^{*_6}$&$ M_{\rm J}$ & $\AU$&$\AU$&&& $J_s^{*_2}$& $M_\odot$ & $*_7$&$*_8$& $*_{9}$  \\
\hline
Source & \multicolumn{9}{|c|}{${\rm Observations}^{*_0}$  }  & \multicolumn{5}{|c|}{${\rm Calculations}^{*_5}$}\\
\hline
Source &  \multicolumn{9}{|c|}{${*_{10}}$  } &E\ref{eq:Jpm}& &$E\ref{eq:etaj}$&$E\ref{eq:etam}$ & \\
\hline
HW Vir &0.485 & 0.142 &0.857& 14.3& 30&4.69&12.8&0.4&0.05&3.16&1        &0.58& 6.8& X\\
\hline
NN Ser &0.535 & 0.111 & 0.93&6.91&2.28&5.38&3.39&0&0.2&0.48& 2          &0.08& 12.2& V\\
\hline
QS Vir & 0.78&  0.43& 1.265& 9.01&56.59&6.32&7.15&0.62&0.92&2.49& 3.5  &0.08& 3.3 &V\\
\hline
RR Cae &0.44 & 0.183 & 1.617& 4.2& &5.3&&0&&0.23& 1                     &0.03& 4.4&V \\
\hline
HS0705 & 0.483& 0.134 &3.464 &31.5 & &3.52 &&0.38&&1.25& 1             &0.25& 4.1& X\\
\hline
HS2231$^{*_{11}}$ & 0.47& 0.075 & 0.789& 13.94 & & 5.16 &&&&0.69& 1                &0.23& 8.8& X\\
\hline
NSVS& 0.46&0.21  & 0.844&2.8& 8&1.9&2.9&0&0.52& 0.38& 1                &0.05& 2.4&V \\
\hline
NY Vir$^{*_{12}}$ &0.459& 0.122 & 0.7585& 2.3 & 2.5& 3.3&5.08&&& 0.22& 1          &0.05& 5.3&V\\
\hline
V471& 0.84& 0.93 & 3.283& 46  & &12.7&&0.26&& 6.2& 3.5                &0.1 &5.7&?\\
\hline
UZ For &0.71 &0.14  & 0.7846& 6.3 & 7.7& 5.9&2.8&0.04&0.05&0.77 & 3.5  &0.07&14.5&V\\
\hline
HU Aqr$^{*_{13}}$  &0.8 & 0.18 & 0.816& 7.1& & 4.3&&0.13&&0.43 & 3.5               &0.03 &12&V \\
\hline
DP Leo$^{*_{11}}$ &1.2 &  0.14& 0.7266& 6.05&  & 8.19&&0.39&&0.55 & 6.5            &0.04 &32.9&V\\
\hline
\end{tabular}
\newline
$*_0$ The sources for the observed data are the papers of \cite{Zorotovic2013} and \cite{Schleicher2014}. For the purpose of this
study we do not address the inaccuracies in measurements (for details see \citealt{Zorotovic2013}).
Cautionary should be applied for the listed properties of some systems:
For HU Aqu other studies specify two planets rather than one, and
the parameters of HW vir, QS vir, and NSVS1425 are in debate, as some studies claim these are unstable
(e.g., \citealt{Horner2012c, Horner2013, Wittenmyer2013, Hinse2014}). \cite{Lohr2014} examine the presence of planets in several of the systems studied here. They find a strong evidence for the existence of two planets in HW~Vir, a less robust evidence for planets in NY~vir, QS~Vir and NSVS~1425, while in HS~2231 they find no support for the presence of a planet.
\newline
$*_1$ Planets masses are $m_{jp} \sin i$, where $j=1,2$ for the two planets and $i$ is the inclination angle of the system.
\newline
$*_2$  $J_s \equiv M_\odot \AU \km \s^{-1}$.
\newline
$*_3$ The initial mass of the primary star of the massive WDs ($M>0.55M_\sun$) is based on \cite{Claeys2014}; their figure 1.
The lighter primaries are assumed to be descendants of solar type stars who terminated their evolution on the RGB (RGBD for RGB descendants).
For NN Ser we take $M_{1,0}=2M_\odot$ \citep{Mustill2013,Schleicher2014}.
\newline
$*_4$ $\eta_{J2}$ and  $\eta_{L2}$ are $\eta_J$ and $\eta_{L}$, respectively, calculated for $a_0=2 \AU$.
\newline
$*_5$ E$k$ stands for equation ($k$).
\newline
$*_6$ $M_{\rm J}$ stands for Jupiter mass.
\newline
$*_7$ Angular momentum transfer efficiency factor.
\newline
$*_8$ Specific angular momentum efficiency factor.
\newline
$*_{9}$ X means that the SG scenario is unlikely, ? means that the SG scenario is borderline and V means that the SG scenario is possible by angular momentum considerations.
\newline
$*_{10}$ More details on specific systems can be found in \cite{Guinan2001,Schreiber2003,Beuermann2010, Kilkenny2011}.
\newline
$*_{11}$ In both HS 2231 and DP Leo systems $M_1$ and $M_2$ are assumed and not derived (see \citealt{Zorotovic2013} and references within), leading to high uncertainties.
In \cite{Schleicher2014} a second options for DP Leo is mentioned (their table 2); we take option 1 for this system (line 12 in table 2 of \citealt{Schleicher2014}).
 The second option with a lower WD mass of $M_1=0.6M_\odot$ is lacking the orbital period. The first option of $M_1=1.2M_\odot$ has $P=0.06\days$ which is consistent with table 2 in \cite{Zorotovic2013} (where the planet is given) and with \cite{Beuermann2011}.
Therefore, we take the first option.
\newline
$*_{12}$  $m_{2p}$ is a very uncertain value for this system (see  \citealt{Zorotovic2013}).
\newline
$*_{13}$ HU Aqr might have two planets with different orbital parameters \citep{Horner2012b, PortegiesZwart2013}.
In this case \cite{PortegiesZwart2013}, who assumes $M_{1,0}=1.6M_\odot$, finds that it is possible that these planets were formed as FG planets.
}
\label{Tab:Table1}
\end{table}

In the $10^{\rm th}$ column we present our calculation of the minimum angular momentum of the planetary system
\begin{equation}
J_{pm} \equiv \sum \mu_{pj} \sqrt{G M_{tj} (1-e^2_{pj}) a_j},
\label{eq:Jpm}
\end{equation}
where the sum is over the planets $j=1$ or $j=1,2$, $\mu_{pj}$ is the reduced mass of planet $j$,
$M_{tj}$ is the total mass inner to the planet including the planet itself, and $e_{pj}$ and $a_j$ are the eccentricity and semi-major axis, respectively, of planet $j$.
It is a minimum angular momentum of the planetary system since the masses of the planets are minimum masses.

In the SG scenario the system starts with the primary and a secondary star. We assume that the secondary star does not change much during the evolution, although it can gain and lose mass. It can gain mass by accretion during, and mostly at the end, of the CE process \citep{Soker2014}, and in CV systems
the secondary loses mass to the primary (UZ For, HU Aqr, DP Leo). \cite{PortegiesZwart2013} studied the stability of the debated two planets around the CV system HU Aqr, and concluded that a FG stable system can exist in such a case.
In the three systems with the strongest constrain (marked X in Table 1) the secondary did not lose
mass and more likely accreted mass. If indeed their initial masses were lower, the angular momentum constrain to be derived below is stronger even.
The initial angular momentum of the stellar binary system is given by
\begin{equation}
J_{b0}=\mu_0  \sqrt{G M_0 a_0 (1-e^2_0)} = 5.9 \left(\frac{\mu_0}{0.13 M_\odot} \right) \left( \frac{M_0}{1.15 M_\odot}
\right)^{1/2} \left( \frac{a_0}{2 \AU} \right)^{1/2}
(1-e^2_0)^{1/2} M_\odot \AU \km \s^{-1},
\label{eq:J0}
\end{equation}
where, $M_0=M_{1,0}+M_2$ is the total mass, $M_{1,0}$ and $M_2$ are the mass of the progenitor and the secondary, respectively,
$\mu_0 = M_{1,0} M_2 / M_0$ is the initial reduced mass, $a_0$ is the initial semi-major axis, and $e_0$ is the initial
eccentricity; we will assume initial circular orbits.

We can consider three types of systems according to the primary present mass.
\newline
(a) Systems that have a primary of $M_1 \la 0.5 M_\odot$. These are descendants of interaction on the RGB.
For these we assume an initial primary mass of $1 M_\odot$, based on the initial-final mass relation given by \cite{Claeys2014}.
We note that \cite{Zorotovic2013} take $M_{1,0}\simeq 1.2 M_\odot$. However, the typical mass of the secondary star in their study is $\sim 0.6 M_\sun$,
while the typical mass of the observed secondary stars is $M_2 \simeq 0.15 M_\odot$.
Therefore, for RGB descendant (RGBD) primaries we take $M_{1,0}= 1 M_\odot$ (see also remark $*_3$ in Table 1).
The case of NN Ser is on the border, and probably the primary reached the early AGB with an initial mass of $\sim 2 M_\odot$ \citep{Mustill2013}.
\newline
(b) Systems where the primary remnant is a massive WD, $0.7<M<0.85 M_\odot$,
that was the core of an AGB star. For these we take
the initial primary mass to be $3.5M_\odot$, based on the relation given by \citealt{Claeys2014}).
\newline
(c) One system, DP Leo, has a very massive WD remnant of $\sim 1.2 M_\sun$, that is likely to be a descendant of a massive AGB star.
This mass is assumed rather than derived accurately (for details see \citealt{Zorotovic2013} and references therein).
For this system we take the initial primary mass to be $6.5M_\odot$.
For such a massive WD, the system avoids the CE phase on the RGB, but in our proposed scenario tidal interaction increases mass-loss and therefore orbital separation
increases before the AGB phase.

Our estimated values of $M_{1,0}$ are listed in Table 1 (column 11). The strongest constrain we derive below
on the SG scenario comes from RGBD. On the RGB the star reaches a maximum
radius of $\sim 0.5-1 \AU$, and tidal interaction can bring a companion from a distance of up to $a_0 \sim 4 R$ \citep{Soker1996}.
We therefore take the initial orbital separation of the primary and secondary to be $a_0=2 \AU$.
This is more or less a maximum initial separation; the average will be lower.

\subsection{The efficiency parameters}
\label{sec:SGparameters}
We now define two efficiency parameters.
The first one is the fraction of initial angular momentum that ends in the planetary system according to the SG scenario, $\eta_J$
\begin{equation}
\eta_J= \frac{J_{pm}}{J_{b\delta}}; \quad  \eta _{J2} \equiv \eta_J(a_0=2AU).
\label{eq:etaj}
\end{equation}
Here, $J_{b\delta}$ is defined as the difference between the initial and final angular momentum according to the SG planet formation scenario:
$J_{b\delta}=J_{b0}-J_{bf}$, where $J_{bf}$ is the final angular momentum of the inner stellar binary system, and $J_{b0}$ is given by equation \ref{eq:J0}.
In many scenarios for planet formation a small fraction of $<0.1$ of the mass in the proto-planetary disk ends up in planets (e.g.,
\citealt{Alibert2005, Alexander2013}). In the SG scenario studied here, even if all the envelope mass ends up in the
proto-planetary disk, we expect that only $\la 0.1$ of the initial disk mass, hence angular momentum, will end up in the planets.
As some envelope mass will be lost in the wind rather through the proto-planetary disk, we expect that a lower fraction, $\eta_J<0.1$,
of the initial angular momentum will end in the planets.

The second efficiency parameter is based on the specific angular momentum. It is defined as the ratio of the angular momentum that ended in the planetary
system per unit mass of the planets, $j_{pm}=J_{pm}/m_p$, where $m_p=m_{1p}+m_{2p}$,
to that of the entire ejected envelope (again, a minimum value is given to the angular momentum of the planetary system).
The value of $j_{pm}$, defined for the observed planets, is about equal to the angular momentum per unit mass of the disk
 from which the planets have been formed.
The envelope mass lost is $M_{\rm env} =M_{1,0}-M_{1}$.
This specific angular momentum efficiency factor is introduced by \cite{Schleicher2014} as $\alpha_L$  where they assumed that $\alpha_L \sim 10$. Namely, an enhancement of an order of magnitude in specific angular momentum between that of the primary envelope and the proto-planetary disk is required. Since \cite{Schleicher2014} calibrated $\alpha_L$ and we calculate it, we use a different notation $\eta_{L}$  which is given by
\begin{equation}
\eta_{L} = \frac{j_{pm}}{(J_{b\delta})/M_{\rm env}} \simeq  \frac{\eta_J}{m_p/M_{\rm env}} ; \quad \eta_{L2}\equiv \eta_L(a_0=2\AU).
\label{eq:etam}
\end{equation}
If, for example, the entire envelope of the primary ends in the disk then $\eta_{L} \simeq 1$.

We can crudely estimate the value $\eta_{L}$ in a simple model where we assume that the proto-planetary disk in the
 SG scenario is formed by mass lost through the second lagrangian point $L_2$.
A detailed calculation that considers the amount of material lost at different radii as the binary systems spirals-in
is beyond the scope of this paper, and includes many uncertainties, like tidal interaction and enhanced mass-loss rate due to binary interaction.
 The ratio of specific angular momentum in the envelope to that lost from $L_2$ is
\begin{equation}
 \eta_{L} \sim \frac {\omega r^2_{L_2}}{\xi \omega R^2_g},
\label{eq:etacal1}
\end{equation}
where $\xi M_{\rm env} R_g^2$ is the moment of inertia of the primary envelope,
and $\omega$ is the angular velocity of both the primary envelope and the binary system,
as we assumed synchronized orbit. $R_g$ is the radius of the primary which is a giant star.
Substituting typical values of the parameters we obtained
\begin{equation}
 \eta_{L} \sim 20 \left( \frac {r_{L_2}}{2 R_g} \right)^2 \left(
\frac {\xi}{0.2} \right)^{-1} .
\label{eq:etacal2}
\end{equation}
{Examining column 13 of Table 1 we find that there is no problem for the SG scenario with regards to the high specific angular momentum in the proto-planetary disk.

\subsection{Challenges to the second generation (SG) scenario}
\label{sec:SGconclusion}
We can divide the systems into three groups according to the constraints from angular momentum considerations.
\begin{enumerate}
\item Systems where $\eta_{J2}< 0.1$ (marked `V' in column 14 in Table 1). In these system the efficiency factors imply that
binary system had sufficient initial angular momentum to account for the formation of a proto-planetary disk and planets,
 under the assumption that not less than $\sim 0.1$ of the disk mass ended up in forming the planets.
\item Systems where $\eta_{J2}\sim 0.1$ (marked `?' in column 14 in Table 1).
In these systems the efficiency factors imply that the SG scenario might be possible, although the transfer of gas
from the proto-planetary disk to the planets must be very efficient.
Other considerations should be taken into account. For example, in the case of NN~Ser dynamical instabilities (e.g., see \citealt{Volschow2014}) might make
the SG scenario more likely than the FG scenario.
Our conclusion regarding the formation mechanism of these planets are consistent with the conclusion of \cite{Schleicher2014} regarding V471.
\item Systems where $\eta_{J2}> 0.2$ (marked `X' in column 14 in Table 1).
In these systems a very large fraction of $>0.2$ of the proto-planetary disk must end up in planets.
This is a large fraction considering the view on planet formation \citep{Alexander2013}.
 Therefore planets will probably not be formed by the SG planet formation scenario in these systems.
 {\cite{Schleicher2014} also found that the predicted planets' masses in these systems are considerably smaller than what is observed. }
\end{enumerate}

~From angular momentum considerations the SG scenario has difficulties with about $25\%$ of the systems,
This assessment is based on present planet formation models that take a fraction
of $<0.1$ of the proto-planetary mass to end up in forming planets \citep{Alexander2013}.
However, none of the systems has a planetary system angular momentum larger than the estimated initial angular momentum of the binary system.
This might be a rescue wheel for the SG scenario.

As can be seen from Table 1 the planets orbital separations around the studied PCEBs ($a<10\AU$) disagree with the predictions
of the disk instability model ($a>20\AU$; see \citealt{Boss2011,Zorotovic2013}). As noted by \cite{Zorotovic2013}  migration \citep{Beuermann2013} and/or scattering
can explain this phenomenon. However, as pointed out by \cite{Zorotovic2013} it seems that SG disks are highly unstable, fragment and tend to form giant planets,
while disks around young single stars and MS binaries only form planets in $\sim 0.1$ of the cases.
These models assume protoplanetary disks similar to the formation of the solar system, which might not hold in PCEBs.
The high frequency of planets around PCEB \citep{Zorotovic2013} and the short time scale of the SG planet formation can result from the enhanced dust to gas mass ratio as claimed by \cite{Zorotovic2013}. However, this enhancement does not exist in He WDs.
{\cite{Zorotovic2013} suggest that enrichment in PCEBs with sdB might occur very close to the tip of the RGB  (for more details see \citealt{Boyer2010}).

\section{THE FIRST GENERATION SCENARIO}
\label{sec:FG}
\subsection{Assumptions and basic processes}
\label{sec:FGassumptions}
To evaluate the FG scenario we calculate the factor $\eta_{\rm ML}$ by which the orbital separations of the planets increase due to mass-loss
\begin{equation}
a_{jp,0}=\eta_{\rm {ML}}a_{jp} =\frac{M_1+M_2+\sum m_{jp}}{M_{1,0}+M_2+\sum m_{jp}}a_{jp},
\label{eq:a_0}
\end{equation}
where $m_{jp}$ is the mass of the planet, and the sum is over the planets j = 1 or j = 1, 2. $a_{jp}$ are the observed planet orbital separations ($j=1,2$), and $a_{jp,0}$ are the calculated initial separation. 
The second equality defines the mass-loss factor $\eta_{\rm ML}$,
and we assume that the mass of the secondary star and planets did not change along the evolution.
The initial orbital separations of the planets calculated by equation (\ref{eq:a_0}) are presented in columns 8 and 9 in Table 2.
The other quantities listed in Table 2 are, according to their columns, as follows.
(1-2) the primary and secondary masses, as in Table 1;  (3) the estimated progenitor mass as as column 11 in Table 1;
(4-7) the planetary masses and period (Table 1);
(10) the resonance between the planets periods;
(11) the initial orbital separation of the stellar binary system assuming a 1:2 resonance between the secondary and the inner plan periods.
\renewcommand{\tabcolsep}{1pt}
\begin{table}
Table 2: Systems - FG Scenario
\newline
\bigskip
\footnotesize{
\begin{tabular}{|l|c|c|c|c|c|c|c||c|c|c|c|}
\hline
\small Column & 1 & 2& 3 & 4 & 5 &6& 7& 8& 9&10& 11  \\
\hline
\small Name &$M_1$ &$ M_2$& $M_{1,0}$ &$m_{1p}$&$m_{2p}$&$P_{1p}$&$P_{2p}$&   $a_{1p,0}^{**_3}$ & $a_{2p,0}^{**_3}$ & $P_{1p,0}:P_{2p,0}^{**2}$&$a_{0}^{**1}$  \\
\hline
units &$M_\odot$&$M_\odot$ & $M_\odot$ &$M_J$&$M_J$&$\yr$&$\yr$& $\AU$&$\AU$&& $\AU$   \\
\hline
Source & \multicolumn{2}{|c|}{${\rm Observations}^{**_0}$  }& {\rm Assumed}& \multicolumn{4}{|c|}{${\rm Observations}^{**_0}$  }  & \multicolumn{3}{|c|}{\rm Calculations}&{\rm  Assumed}   \\
\hline
HW Vir &0.485&0.142& 1&14.3&30&12.7&55&2.60&7.24&2:9&1.63 \\
\hline
NN Ser  &0.535&0.111& 2&6.91&2.28&15.5&7.75&1.65&1.05&(1.05)3:2 &0.66 \\
\hline
QS Vir  &0.78&0.43& 3.5&9.01&56.59&14.4&16.99&1.96&2.28& 4:5 &1.23\\
\hline
RR Cae  &0.44&0.183& 1&4.2&&11.9&&2.80&&&1.76\\
\hline
HS0705  &0.483&0.134& 1&31.5&&8.41&&1.96&&&1.22\\
\hline
HS2231  &0.47&0.075& 1&13.94&&15.7&&2.65&&&1.66\\
\hline
NSVS    &0.46&0.21& 1&2.8&8&3.49&6.86&1.05&1.62&(1.06)1:2&0.66\\
\hline
NY Vir  &0.459&0.122& 1&2.3&2.5&7.9&15&1.71&2.64&(1.05)1:2&1.08\\
\hline
V471  &0.84&0.93& 3.5&46&&33.2&&5.15&&&3.23\\
\hline
UZ For   &0.71&0.14& 3.5&6.3&7.7&16&5.25&1.39&0.66& 3:1&0.42 \\
\hline
HU Aqr   &0.8&0.18& 3.5&7.1&&9&&1.15&&&0.725\\
\hline
DP Leo   &1.2&0.14& 6.5&6.05&&28.01&&1.66&&&1.04\\
\hline
\end{tabular}
\footnotesize
\newline
$**0$ The sources for the observed data are the papers of \cite{Zorotovic2013} and \cite{Schleicher2014}.  For the purpose of this
study we do not address the inaccuracies in measurements (for details see \citealt{Zorotovic2013}).
\newline
$**1$ $a_{0}$ is defined as the initial orbital separation of the binary (primary and secondary). The initial orbital separation of the binary is calculated under the assumption of a stable resonance of 1:2 of the secondary ($M_2$) with the closest planet.
Under this assumption $a_{0}$ is smaller for most systems than the assumed $a_{0}=2\AU$ in section \ref{sec:SG}.
Taking a lower value for $a_{0}$ in section \ref{sec:SG} will result in less available angular momentum for planet formation,
and hence makes the SG scenario less likely.
\newline
$**2$ The resonance are calculated as $\frac{P_{1p,0}}{P_{2p,0}}$. If the deviation from the written two integers ratio is
larger than $2 \%$ we indicate the factor of deviation in the parenthesis.
\newline
$**3$ $a_{1p,0}$ and $a_{2p,0}$ are defined as the calculated initial orbital separation of the planet according to the FG scenario (Eq. \ref{eq:a_0}).
}
\end{table}

We further note that there are resonances between the orbital periods of planets (column 10 in Table 2).
Some of these resonances have been reported previously (e.g., NN~Ser, \citealt{Beuermann2013}).
To be able to proceed, we make the assumption that in the initial system, before the primary left the MS, the secondary star
and the inner planet were in a resonance and we take it to be a 1:2 resonance.
This might make the system stable, and allow us to build the scenario discussed below.
{{This assumed 1:2 resonance is a conservative one for the FG scenario. Recent paper by \cite{Zhang2014} state that a significant fraction of planets are in a 1:2 resonance. As we suggest below if the secondary star formed as planets forms, it is likely that the secondary and the inner planet were formed in a resonance.
Any stronger resonance, such as a 2:3 resonance as Neptune and Pluto have, implies
that the initial orbit of the secondary star is closer to the inner planet, hence at a larger orbital separation from the primary star.
Allowing the secondary star to have a larger initial orbital separation increases the parameter space for the allowed progenitors,
and make the FG scenario more probable. }}

In systems where the initial secondary orbital separation is $\geq 1 \AU$, we can allow according to our
 proposed scenario the secondary to have a lower initial orbital separation, e.g., to be 0.63 times the orbital separation given
  in Table 2 and have a 1:4 resonance with the inner planet.
The results of \cite{Mustill2013} show that some initial configurations make the binary system with planets very long-live
(the white squares in their figure 3).
Such configurations are systems in a resonance. We conservatively assume a 1:2 resonance between the binary system and the inner planet.
The presence of an outer planet might somewhat shift the stable configuration from this exact 1:2 resonance.

\subsection{The first generation scenario}
\label{sec:FGscenario}
The proposed scenario for all systems is based on a strong tidal interaction during the RGB phase, for the RGBD systems
(i.e., primaries that are descendants of RGB stars: He WD, horizontal branch stars, or  CO WDs that are descendants of HB stars).
However, an early CE phase is avoided because the secondary star is massive enough to bring the
system to synchronization between the rotation of the primary envelope and the orbital motion of the secondary star.
A CE phase occurs only later, on the late RGB for the RGBD progenitors, and on the late AGB phase for the CO (or ONeMg) WD progenitors.

We consider the following scenario for the RGBD systems. The progenitors of these stars, assumed to be $M_{1,0} \sim 1 M_\sun$ on
the MS, expand up to $\sim 200 R_\odot \simeq 1 \AU$ on their RGB (e.g., \citealt{IbenTutukov1985}). From our estimated initial
orbital separation of the secondary stars, last column in Table 2, we see that a strong tidal interaction
takes place during the RGB phase of the primary. The secondary brings the primary envelope to synchronization with the orbital motion, and
a CE phase is avoided as interaction starts. The primary RGB star loses mass and expands as its core's mass grows, and eventually a CE phase
 is unavoidable. However, by this stage the envelope mass is lower than its initial value, and the secondary star survives the CE phase despite its low mass.
The planets move outward due to the mass-loss process. The mass-loss process during the CE phase can be on a time scale
not much longer than the  planets' orbital period, which can explain the non-zero eccentricity in some cases.

Stars starting with a mass of $M_{1,0} \ga 2.3 M_\sun$ on the MS expand much less during their
RGB phase (e.g., \citealt{IbenTutukov1985}). Therefore, the scenario plotted above for the RGBD systems occurs also for the more massive WD systems,
but during the AGB phase of the primary star. These estimations are in accordance with \cite{Zorotovic2013} who noted that the range in orbital separation in the initial binary is
$\sim 0.5\AU$ to $\sim 2.5\AU$ depending on the final mass of the primary.

In DP Leo the progenitor of the massive WD was most likely a massive star of $\ga 6 M_\sun$ on the MS.
How did the a secondary star of mass $M_2=0.14 M_\odot$ survived the CE phase in a massive
envelope of such a primary star?  If the secondary star was formed as a planet does, as we suggest in the present study,
it is plausible that an even inner body, a massive planet or a brown dwarf, was presence in the system.
This inner body entered the envelope of the primary at an earlier stage and did not survive its CE phase.
But as it spiraled-in and collided with the primary core it ejected a large fraction of the primary envelope.
This enabled the secondary star to survive its CE phase \citep{BearSoker2011}, now taking place in a much lighter envelope.

In a very recent paper \cite{Bruch2014} suggests that a planet of  $9.5 M_{\rm J}$
orbits the cataclysmic variable V893 Sco at an orbital separation of $4.5\AU$.
Interestingly, the stellar secondary mass in this systems is $m_2 \sim 0.13-0.18M_\odot$,
very similar to most secondary masses in the systems studied here. The low mass of the secondary stars in most studied PCEBs deserve some attention.
\cite{Zorotovic2013} performed a binary population study of these PCEBs to characterize their main sequence binary progenitors.
The average secondary mass in their population synthesis study was $\sim 0.6 M_\sun$, much larger than the value in most systems studied here.
This large discrepancy might hint that the secondary was not formed as a star forms, but rather like massive planets form.

Another aspect of the low-mass secondary stars are the way they interact with the evolving primary star.
They are not sufficiently massive to eject the primary envelope before they spiral deeply in, but they are massive enough to bring the envelope to synchronization before they enter the CE phase.
After the primary envelope spin is synchronized with the orbital motion, a Darwin instability might occur.
The instability sets in when the moment of inertia of the primary envelope $\xi M_{\rm env} R_g^2$ exceeds third of the orbital moment of inertia, $\sim M_2 a^2$, where $a$ is
taken at synchronization. The condition for instability to occur reads
\begin{equation}
M_2 \leq 0.3\left(\frac{\xi}{0.2}\right)\left(\frac{M_{env}}{0.5M_\odot}\right)\left(\frac{R_g}{a}\right)^2 M_\odot
\end{equation}
For orbital separation at synchronization of $R_g < a \la 2R_g$ the condition for instability to occur is $M_2 \la 0.075-0.3 M_\sun$.  This shows that the secondary stars in most system
will enter the envelope due to Darwin instability. This, we argue, what makes these systems unique.

Finally we note that even in the FG scenario the already existing planets can accrete some of the mass lost by the stellar binary system (e.g., \citealt{Perets2010,Beuermann2013,Zorotovic2013,Schleicher2014}).
In this hybrid scenario the low-mass planets can be treated as "seeds" that can accrete more mass from the disk.
The mass accretion can reduce orbital separation of planets. This hybrid scenario can bridge between the FG scenario and the SG scenario. However, in most systems we find that the FG scenario can work without the aid of mass accretion.

\section{SUMMARY}
\label{sec:Summary}
We compared the the second generation (SG) planet formation scenario with the first generation
(FG) planet formation scenario for twelve post common envelope binary (PCEB) systems that host planets.
We concentrated on constraints imposed from angular momentum evolution. In comparing the FG with SG scenarios one should examine what is more likely, and
whether one of the possibilities can be ruled out. Doing that, we reach the following conclusions.
\begin{enumerate}
    \item When considering the SG scenario we find that in none of the PCEBs the planetary system angular
     momentum is larger than the estimated initial stellar binary angular momentum.
    However, in three systems (marked X in the last column of Table 1), our calculated efficiency of angular momentum transfer from the stellar binary system
    to the planetary systems is extremely high, $\eta_{J2} > 0.2$. It implies that more than $20 \%$ of the proto-planetary disk mass
    in these three systems ended up in planets. This is a very high efficiency \citep{Alexander2013} that imposes strong challenges to the SG scenario.
    \item We found that all planets in the twelve systems can be FG planets.
    However, when the primary was a main sequence (MS) star, the planets were closer to the center; they
    moved outward during evolution due to mass-loss. The secondary star, on the other hand, had to be at an orbital separation of $\ga 0.5 \AU$
    to allow the primary to develop a massive core.
    Hence, the orbits of the secondary star and of the inner planet were quite close to each other in many systems, according to the FG scenario.
    To maintain such close orbits requires that in such systems the orbit of the stellar companion and the inner planet were in a resonance.
    A support to the presence of a resonance is the resonance that exists today (e.g., \citealt{Beuermann2010, Beuermann2013, Horner2012a, Mustill2013, Volschow2014})
   between the two planets in many of the systems (columns 6 and 7 in Table 2).
   For that, in estimating the initial stellar binary orbital separation we assumed a resonance of 1:2 between the secondary and the inner planet
   (column 11 in Table 2). This resonance between the inner planet and the secondary can stabilize the system and support the FG scenario.
    Resonances similar to the ones assumed here are present in
    many exoplanet systems between the planets (e.g.,\citealt{Marcy2005, Tinney2006, Raymond2008, Gozdziewski2014,Zhang2014}).
    Here we predict that careful monitoring over several years of binary systems where the secondary is a very low-mass  main sequence star,
     $M_2 \simeq 0.1-0.2 M_\odot$, with an orbital period of several months to several years will reveal that many of them have
     planetary systems in resonance with the binary system.
    \item Following the previous conclusion, in the FG scenario it is quite likely that the secondary star was
    formed as a planet, but accreted mass from the proto-planetary disk of the young star, as it is not easy to form planets around binary stars (e.g., \citealt{Linesetal2014}).
\end{enumerate}

Overall, we consider the FG scenario more likely, at least from angular momentum considerations.
However, \cite{Schleicher2014} point out that there are probably two populations, where some systems form through the FG scenario and some by the SG scenario.
In future studies other aspects (e.g., stability, typical
lifetime) should be incorporated in the model for a better understanding of the overall picture.
In some of the systems FG planets might accrete mass.
Such a process can enrich the outcome of PCEB evolution.
Additionally, the existence of a low-mass body closer even to the center than the stellar companion in the progenitor system
can help the stellar binary survive the evolution. It does so by increasing the mass lost from the primary prior to the entrance
of the secondary star to the primary envelope.

Many other PCEB systems are known. Currently many of them are not known to host planets \citep{Schreiber2003,Zorotovic2013,Schleicher2014}.
These systems might be good candidates for planet observations.
Furthermore, \cite{Schaffenroth2014} recently discovered a PCEB where the primary has a mass of $M_1=0.47M_\odot$ and the companion is a brown dwarf of mass $M_2=0.064M_\odot$.
This brown dwarf survived the CE phase and probably triggered the mass-loss process that transformed the progenitor of the primary to a sdB star.
Today observations point out that the primary masses of PCEBs with planets exclude $M_1\sim 0.6M_\odot$.
The lack of this typical range in mass in PCEBs where planets have been found is probably the result of the evolutionary process.
Further research on this topic is required.

We would like to thank Dominik Schleicher for helpful comments.
We thank Alexander Mustill for comments that strengthen our conclusions supporting the FG scenario.
We thank an anonymous referee for comments that improved our manuscript.
This research was supported by the Asher Fund for
Space Research at the Technion and the US - Israel Binational Science Foundation.

\end{document}